\documentclass[11pt]{article}%
\usepackage[margin=1.15in]{geometry}
\usepackage{amsmath}
\usepackage{amsfonts}
\usepackage{amssymb}
\usepackage{graphicx}
\usepackage{hyperref}%
\begin{document}

\title{Anthropic reasoning and typicality in multiverse cosmology and string
theory\thanks{Thanks to Yuri Balashov, Gordon Belot, Rob Caldwell,\ Marcelo
Gleiser, Brad Monton, Ken Olum, Jim Peebles, Lee Smolin, and Alex Vilenkin for
helpful discussions and comments on an earlier draft.}}
\author{\bigskip Steven Weinstein\thanks{Email: sw@uwaterloo.ca}\\{\small Perimeter Institute for Theoretical Physics, 31 Caroline St, Waterloo,
ON \ N2L 2Y5 Canada}\\{\small Dept. of Philosophy, University of Waterloo, Waterloo, ON N2L\ 3G1
\ Canada}\\{\small Dept. of Physics, University of Waterloo, Waterloo, ON N2L\ 3G1
Canada}}
\date{November 25, 2005}
\maketitle

\begin{abstract}
Anthropic arguments in multiverse cosmology and string theory rely on the weak
anthropic principle (WAP). We show that the principle is fundamentally
ambiguous. It can be formulated in one of two ways, which we refer to as
WAP$_{1}$ and WAP$_{2}$. We show that WAP$_{2}$, the version most commonly
used in anthropic reasoning, makes no physical predictions unless supplemented
by a further assumption of \textquotedblleft typicality\textquotedblright, and
we argue that this assumption is both misguided and unjustified. WAP$_{1}$,
however, requires no such supplementation; it directly implies that any theory
that assigns a non-zero probability to our universe predicts that \emph{we}
will observe our universe with probability one. We argue, therefore, that
WAP$_{1}$ is preferable, and note that it has the benefit of avoiding the
inductive overreach characteristic of much anthropic reasoning.

\end{abstract}

%\pacs{11.25.-w, 11.25.Mj, 98.80.-k, 98.80.Cq}

\section{Introduction}

Over the last twenty years or so, inflationary cosmologists have been toying
with theoretical models that postulate the existence of a multiverse, a set of
quasi-universes (henceforth \textquotedblleft universes\textquotedblright)
which are more-or-less causally disjoint \cite{Vil83},\cite{Linde86}. String
theorists, who once hoped to predict a unique supersymmetric extension to the
standard model, now find themselves contemplating a similar scenario, with
perhaps $10^{500}$ or more metastable low-energy vacua (the \textquotedblleft
landscape\textquotedblright) realized via inflationary mechanisms as
effectively distinct universes \cite{BP00},\cite{Sus03},\cite{KKL03}%
,\cite{FKRS05}. The vast majority of the universes in these scenarios look
nothing like our universe, the values of the fundamental physical parameters
(e.g., the cosmological constant) differing markedly from the values we
observe. A significant number of physicists have understood this situation ---
in particular, the fact that universes like ours are atypical --- to be
problematic\ because, it is felt, it reveals a lack of explanatory or
predictive power. Others seem more concerned with our inability to either
confirm or falsify such theories. It is in addressing these concerns that
anthropic reasoning enters the picture.

Anthropic \emph{reasoning} is reasoning based on the use of the (weak)
anthropic \emph{principle}, articulated by Carter as

\begin{quote}
\textbf{WAP}: \textquotedblleft What we can expect to observe must be
restricted by the conditions necessary for our presence as
observers.\textquotedblright\ (\cite{Car74}, p. 291).
\end{quote}

\noindent This seems unobjectionable; in fact, it is a tautology. As such, one
wonders how it can do any explanatory or methodological work, much less be the
object of heated disagreement. We will argue that the anthropic principle, as
stated by Carter, contains what Bohr might have called an \textquotedblleft
essential ambiguity\textquotedblright(\cite{Bohr35}), and that the problematic
aspects of anthropic reasoning stem from the use of a form of the principle
which requires recourse to an \emph{additional} and unsupported assumption of
\textquotedblleft typicality\textquotedblright\ in order to make contact with
observation. Though the crucial importance of the \textquotedblleft principle
of mediocrity\textquotedblright\ which encodes the assumption of typicality is
understood by some, its connection to a particular form of the WAP and its
status as a distinct and unargued-for assumption seems not to be well understood.

\section{Anthropic arguments}

\subsection{General form}

The general strategy of an anthropic argument is as follows. Begin with some
mulitiverse hypothesis, a hypothesis which gives:

\begin{itemize}
\item a set of possible universes parametrized by the values taken by relevant
physical parameters such as the cosmological constant, the dark matter
density, etc., and

\item a probability distribution describing the relative frequency of
occurrence for these universes in the (generally infinite) ensemble.
\end{itemize}

\noindent Now, in accordance with WAP, restrict attention to the subset of
universes that support the existence of observers, since these are the only
universes we could hope to observe, and consider the distribution of the
parameters over this \textquotedblleft anthropic\textquotedblright\ subset.
These are the universes it is possible to observe, and so the predictions for
what we can expect to observe should be based on this subset.

Given the restriction to the anthropic subset, one has a new probability
distribution for each parameter reflecting the relative frequency with which
the various values of the parameter occur in the anthropic subset. If the
parameters take on a continuous set of values, or even a large but finite
number, the probability of any particular value will be absurdly small, so
that no particular outcome is to be expected.\ What, then, does anthropic
reasoning predict?\ 

It is here that one must introduce an assumption of typicality, such as
Vilenkin's \textquotedblleft principle of mediocrity,\textquotedblright\ which
he describes as \textquotedblleft the assumption that we are typical among the
observers in the universe.\textquotedblright\ He elaborates,

\begin{quote}
Quantitatively, this can be expressed as the expectation that we should find
ourselves, say, within the $95\%$ range of the distribution. This can be
regarded as a prediction at a 95\% confidence level. (\cite{Vil04}, p. 2)
\end{quote}

\noindent Equipped with this further assumption, one can infer that the
observed values of the parameters will\emph{\ }lie within the $95\%$
range.\footnote{But note that the notion of typicality, of lying within some
confidence interval, is not well-defined for all probability distributions.
\ For example, a flat distribution has no \textquotedblleft
typical\textquotedblright\ values. \ What, for example, is a typical outcome
of the roll of a fair die?} In short, if we assume we are typical (i.e., if we
apply the principle of mediocrity), we get a range of predicted outcomes which
is a proper subset of the anthropic subset.

From here, the anthropic reasoner will generally conclude that the observation
of an (anthropically) typical value by us means that the theory has
successfully \emph{explained} the value (since we are a generic
\emph{prediction} of the theory). Furthermore, she will judge that the
observation of a typical value offers inductive \emph{support} to the theory.

Similarly, observation of an \emph{a}typical value is judged to constitute a
disconfirmation, even falsification, of the theory \cite{Rees03},\cite{Dav04}.
However, from a logical standpoint, all that has been shown\ by a failure to
observe an anthropically typical value is that the \emph{conjunction} of the
theory and the principle of mediocrity is inadequate. One might just as well
impugn the principle of mediocrity rather than the theory. The following
example brings this out.

\subsection{Example:\ the Googolverse}

Consider a multiverse theory which postulates the existence of a
\textquotedblleft Googolverse\textquotedblright, an infinite ensemble made up
of $10^{100}$ (a googol) different sorts of universe, where \textquotedblleft
different sorts\textquotedblright\ means that the universes are characterized
by different values of one or more physical parameters. The postulated theory
also provides a probability distribution over the parameters, so that one can
talk about \textquotedblleft typical\textquotedblright\ and \textquotedblleft
atypical\textquotedblright\ values. Suppose, finally, that the values
\emph{we} observe are atypical, lying outside the $95\%$ confidence interval.
In such a situation, we might turn to the anthropic principle in order to
assess the theory.

Suppose that, upon doing an anthropic analysis of this multiverse, we discover
that $10^{5}$ of the $10^{100}$ sorts of universe support observers. In
accordance with the anthropic principle, we reason that these are the only
universes we could possibly observe, and we go on to ask whether what we
observe --- the value of the cosmological constant, the electron mass, and so
on --- is typical of this observer-supporting, \textquotedblleft
anthropic\textquotedblright\ subset. In other words, would a generic observer
expect to see something like what we see? The anthropic reasoner says that if
the answer is no, then our theory should be regarded as an explanatory
failure, and our observation of the atypical values should count against
acceptance of the theory.

So far, so good... perhaps. But suppose we analyze the observer-supporting
subset further, and discover that there are really only two general sorts of
observers, Humans and Aliens, and that Aliens are by far the dominant form of
life in the observer-supporting subset. In fact, we discover that only $17$ of
the $10^{5}$ sorts of universes support the existence of Humans (Aliens are
apparently far more adaptable), and that our probability distribution tells us
that these account for only $0.1\%$ (i.e., $100$) of the sorts of universe in
the observer-supporting subset. Statistical analysis reveals that of this
Human-accommodating subset of $17$ sorts of universe, the values we observe in
our universe, one of the $17$, are entirely typical.

In short, we come out looking like an atypical member of the Googolverse, an
atypical member of the observer-supporting subset, and a typical member of the
Human-supporting subset. What to conclude? Most anthropic reasoning seeks
typicality (or \textquotedblleft mediocrity\textquotedblright) amongst the
broadest possible subset of observers, judging that \textquotedblleft it is
prudent to condition probabilities, not on a detailed description of `us', but
on the weakest condition consistent with `us' that plausibly provides useful
results.\textquotedblright\ (\cite{Har04}, p. 5). Thus most anthropic
reasoners would presumably conclude that the observed values are not
\emph{explained} by the Googolverse hypothesis, and \emph{a fortiori} that
these values do not \emph{support} the hypothesis. On the other hand, the
typical member of the small subset of the human race who is not completely
flummoxed by the idea of a multiverse might well disagree, noting that the
Googolverse hypothesis succeeds very well in explaining what \emph{we}
observe, since \emph{we} are Humans, not Aliens.

\section{The Essential Ambiguity}

The Googolverse example is of course a caricature, but it highlights the way
in which the particular choice of a reference class of observers can make an
enormous difference to the outcome of an anthropic argument. It does so via
the additional requirement of typicality with respect to the reference
class.\footnote{See \cite{Bos02} and \cite{Smo04} for further discussion of
the problem of the reference class.} Given the seemingly uncontroversial
nature of the WAP, it is surprising that its application can be so
contentious. In this section, we will see that this is a result of an
ambiguity in the principle itself.

Consider once again the Weak Anthropic Principle, \textquotedblleft What we
can expect to observe must be restricted by the conditions necessary for our
presence as observers.\textquotedblright\ How are we to understand the phrase
\textquotedblleft our presence as observers\textquotedblright? Are we to
understand it as talking about \emph{our} presence, or the presence of
observers, \emph{in general}? The difference is pivotal.

Given that we \emph{are }observers, the conditions necessary for our presence
as observers are no different from the conditions necessary for our presence
\emph{simpliciter}, and we might reformulate the WAP accordingly:

\begin{quote}
\textbf{WAP}$_{1}$: \textquotedblleft What we can expect to observe must be
restricted by the conditions necessary for our presence.\textquotedblright
\end{quote}

\noindent This is an uncontroversial claim, since \textquotedblleft
we\textquotedblright\ can only observe the properties of worlds that allow our
presence. So for example, we cannot observe a world or a universe in which we
failed to evolve, even if that universe has earth-like planets and other
DNA-based life forms.\footnote{Interpreting the weak anthropic principle in
this way more or less implies the \textquotedblleft top-down\textquotedblright%
\ approach of \cite{AT05}.}

Clearly, most proponents of anthropic reasoning understand Carter's principle
in a different way, interpreting WAP as,

\begin{quote}
\textbf{WAP}$_{2}$: \textquotedblleft What we can expect to observe must be
restricted by the conditions necessary for the presence of
observers.\textquotedblright
\end{quote}

\noindent Application of WAP$_{2}$ requires one to establish what
\textquotedblleft observers\textquotedblright\ are, and then to identify the
conditions necessary for their presence. The constraint is much looser than
the constraint imposed by WAP$_{1}$, in that it admits universes which do not
contain \textquotedblleft us\textquotedblright, do not contain human beings at
all. Furthermore, although WAP$_{2}$ tells us that what we can expect to
observe is \emph{restricted} by the conditions necessary for observers, it
does not tell us anything about the likelihood of \emph{our} observing
particular conditions, even though it does instruct us to calculate the
probability distribution of the parameters over the sub-ensemble of universes
that support the existence of observers. In order to extract concrete
predictions from WAP$_{2}$, it must be \emph{supplemented} by the principle of
mediocrity (or something similar), which stipulates that we are \emph{typical
}observers. But of course we need \emph{not} be, or we may be typical of one
class of observers and atypical of another (as in the Googolverse example
above). \ 

Thus any application of WAP$_{2}$ requires us to choose a particular reference
class with respect to which we choose to assume typicality. \ \ As noted in
the previous section, proponents of anthropic reasoning normally suggest that
we condition on the \emph{broadest} possible reference class of observers.
\ \ However, this is at odds with ordinary practices of statistical inference.
\ When one is attempting to account for selection effects (which is after all
the entire \emph{point} of anthropic reasoning), one does so by conditioning
on as \emph{detailed} a description as possible. \ So, for example, if we know
that the cosmological constant $\Lambda=0.75\pm.0.05$, then we should demand
typicality with respect to this, for this is what we should expect to see,
given what we know about ourselves. \ If we know that $\Lambda=0.75$
\emph{exactly}, then we should condition on this. If one conditions on precise
values for all cosmological parameters, then there is no need at all for a
principle of mediocrity --- no need to assume that we are typical members of
some larger ensemble. \ Taking selection effects as seriously as possible is
thus equivalent to appealing to WAP$_{1}$ as a principle of inference.

Taking selection effects seriously does \emph{not} mean that we are in any way
ignoring the usual role of typicality in statistical reasoning. \ Appeals to
the notion of statistical significance presuppose, not that the
\emph{observer} is randomly chosen (i.e., \textquotedblleft
typical\textquotedblright), but that the \emph{observations} made by a given
observer are random. \ To the extent that this observer does not have access
to the entire ensemble under scrutiny, selection effects, anthropic or
otherwise, must be taken into account. \ Thus we filter out the effects of the
galaxy and other known effects (e.g., the dipole anisotropy) from our
CMB\ observations, and only then require that what remains lie within, say,
the 95\% confidence interval of the statistics generated by the theory we are
interested in confirming.

\section{Conclusion}

WAP$_{1}$ and WAP$_{2}$ are both interpretations of the weak anthropic
principle, and both are incontrovertibly true. Why, then, have we suggested
that one should use WAP$_{1}$ rather than WAP$_{2}$? Because WAP$_{1}$
directly yields testable predictions, while WAP$_{2}$ requires one to both
identify a particular class of observers and apply the principle of mediocrity
with respect to the class chosen. To be sure, WAP$_{1}$ makes the
simultaneously trivial and strong claim that if the theory assigns a
non-vanishing probability to the parameter values we do observe, then that is
what \emph{we} should \emph{expect} to observe.\ If on the other hand the
theory assigns a probability of zero to the values that we observe, then of
course the theory is ruled out.

With respect to theory confirmation, one should certainly not say that the
fact that the universe we observe is predicted with some probability (however
small or large) serves as a confirmation of the theory. \ But then, why would
anyone expect that one could reason inductively from a single data point in
the first place?

\end{document}